\documentstyle[preprint,prl,aps]{revtex}

\begin{document}
\draft
\hyphenation{following}
%
%
\title{Magnetocrystalline Anisotropy Energy
of a Transition Metal Monolayer:
A Non-perturbative Theory}
\author{T. H. Moos, W. H\"ubner, and K. H. Bennemann}
\address{Institute for Theoretical Physics, Freie Universit\"at
Berlin, Arnimallee 14, D-14195 Berlin, Germany}
\date{February 24, 1995}
\maketitle
\begin{abstract}
The magnetocrystalline anisotropy energy $E_{anis}$ for a monolayer of
Fe and Ni is determined using a fully convergent tight-binding calculation
including $s$-$d$ hybridization.
The spin-orbit interaction $\lambda_{so}$ is treated non-perturbatively.
Remarkably, we find $E_{anis}\propto\lambda_{so}^2$ and important
contributions to $E_{anis}$ due to the lifting of degeneracies near the
Fermi-level. This is supported by
the calculated decrease of the anisotropy energy with increasing
temperature on a scale of several hundred K.
Our results clarify the present debate on the origin of $E_{anis}$.
\end{abstract}
Despite increasing theoretical effort, the origin of
magnetic anisotropy in ferromagnetic transition metals has not
been clearly identified.
Actually, present theoretical analyses are
controversial~\cite{wan93a,kel93}.
It remains to determine clearly how the magnetocrystalline anisotropy energy
$E_{anis}$ depends on the spin-orbit
coupling (SOC) strength $\lambda_{so}$ and whether the level crossings at
the Fermi-level $E_F$ play an important role for $E_{anis}$ or
alternatively whether they have
to be excluded as stated by Wang {\em et al.}~\cite{wan93a}. It is important
to present calculations without using the state-tracking method by Wang
{\em et al.}~\cite{wan93a},
the validity of which was disputed by Daalderop {\em et al.}~\cite{kel93}.
Due to increased recent experimental activity
on magnetic anisotropy at surfaces, interfaces and
thin films~\cite{expts,exptsni}, there is need of
a theory without too many restrictive assumptions
in order to get a systematic understanding of $E_{anis}$.

In this paper, we present a theory
treating $\lambda_{so}$ non-perturbatively and determining
the electronic bandstructure within the combined interpolation
scheme~\cite{hod66}.
We investigate a
simple quadratic Fe and Ni monolayer epitaxially grown on the Cu(001) surface
and neglect further interactions with the substrate.
Our calculation permits us to identify clearly the contributions to $E_{anis}$
resulting from level-shifts and lifting of degeneracies near $E_F$.
In particular, the latter give important contibutions to $E_{anis}$.
We find for these that generally $E_{anis}\propto\lambda_{so}^2$ is valid.
The degeneracies do not only appear at high symmetry points of the
Brillouin zone (BZ)~\cite{kel93} but also along lines in {\bf k}-space.
It is very important that we obtain convergent results for $E_{anis}$
without applying state-tracking or excluding surface-pair
coupling as was necessary in previous analyses by
Wang {\em et al.}~\cite{wan93a}.
Furthermore, we find the characteristic scale for the temperature
dependence of the magnetic anisotropy to be $\lambda_{so}$, rather than
the bandwidth, again indicating the significance of the lifting of
degeneracies at $E_F$ by $\lambda_{so}$.

We start our theory by defining the magnetic anisotropy energy as
\[
E_{anis}(n):=E_{tot}(\theta =0;n)-E_{tot}(\theta =\pi /2,\phi _0;n),
\]
where $E_{tot}(\theta ,\phi ;n)$ is the ground-state energy
per atom of the monolayer with a total of $n$ $3d$- and $4s$-electrons
per atom and $N$ atoms
and is given by
\[
E_{tot}(\theta,\phi;n)=\frac{1}{N}
\sum_{m,{\bf k}} E_{m{\bf k}}(\theta,\phi)
 f_0\left(E_F(n)-E_{m{\bf k}}(\theta,\phi)\right).
\]
The angles $\theta $ and $\phi$ denote the direction of the magnetization
$\hat{\zeta}$~\cite{angles}.
$f_0(\Delta E)$ is the Fermi-function at zero temperature and
$E_F(n)$ is the Fermi-energy, which, for a given bandfilling
$n$, is determined self-consistently by
\[
n=\frac{1}{N}
\sum_{m,{\bf k}}
 f_0\left(E_F(n)-E_{m{\bf k}}(\theta,\phi)\right).
\]
$E_{m{\bf k}}(\theta,\phi)$ is the $m$-th eigenvalue with
crystal momentum ${\bf k}$ and magnetization along $(\theta,\phi)$
of the Hamiltonian
\[
H=H_{s}+H_{d}+H_{sd}+H_{so}.
\]
$H_s$ refers to the $s$-electrons which are described by a set of
properly symmetrized plane waves.
$H_{d}$ refers to the $d$-electrons treated by spin-polarized
tight-binding approximation using the Fletcher-Wohlfahrt
parametrization~\cite{fle52,sla54} adapted to the monolayer.
$H_{sd}$ denotes $s$-$d$ hybridization (in the conventional form of
the combined interpolation scheme~\cite{hod66}) and
$H_{so}=\lambda_{so} {\bf l}\cdot{\bf s}$ is the SOC
between the $d$-electrons on the same site in the usual approximation with
the SOC parameter $\lambda_{so}$.
$H_{so}$ is a matrix function of the magnetization direction $\hat{\zeta}$.
It can be expressed in terms of the orbital momentum operators
$l_{\xi}$, $l_{\eta}$ and $l_{\zeta}$ with respect to the rotated frame
$(\hat{\xi},\hat{\eta},\hat{\zeta})$
as~\cite{tak76,bru89}
\begin{equation}
\label{hso}
H_{so}=\left( \begin{array}{cc}
    H_{so}^{\uparrow\uparrow} &
    H_{so}^{\uparrow\downarrow} \\
    H_{so}^{\downarrow\uparrow} &
    H_{so}^{\downarrow\downarrow}
            \end{array} \right)
     =\frac{\lambda_{so}}{2}
           \left( \begin{array}{cc} l_{\zeta} & l_{\xi}-il_{\eta}
              \\ l_{\xi}+il_{\eta} & -l_{\zeta} \end {array} \right) .
\end{equation}

For the numerical analysis,
we employ for the $d$-electrons a basis of 10 two-dimensional Bloch-functions
$\psi_i$, $i=1,...,10$,
for each possible crystal momentum ${\bf k}$,
constructed from atomic 3$d$-wave functions
together with the spin eigenstates $\left|\uparrow\right>$ and
$\left|\downarrow\right>$
of the Pauli matrix $\sigma_{\zeta}$,
where $\hat{\zeta}$ is the
spin quantization axis.
Only nearest neighbors (4 in a simple-quadratic monolayer) are
considered.
To obtain accurate parameters, we perform a fit to the full-potential
LMTO calculation for a free-standing Fe-monolayer by
Pustogowa {\em et al.}~\cite{pus95} and to the LAPW calculation for a
Ni-monolayer by Jepsen {\em et al.}~\cite{jep82}.
The $s$- and $d$- bandwidths and hybridization parameters are then
scaled to take into account the Cu surface lattice constant of
2.56~{\AA}~\cite{har80}.
The complete BZ summation over
${\bf k}$ is performed as a weighted summation over the irreducible part of the
BZ ($1/8$ of the quadratic BZ in the nonmagnetic case or for perpendicular
magnetization and $1/4$ for in-plane magnetization due to reduced
symmetry). To achieve convergence, about 1000 or 2000 points in the
irreducible part of the BZ are sufficient. Note, we do not have to exclude any
parts of the BZ to obtain convergence, unlike Wang
{\em et al.}~\cite{wan93a}.

In Fig.~\ref{fig1} we present results for $E_{anis}$
as a function of the
bandfilling $n$, in order to demonstrate the correspondence between
electronic structure and magnetic anistotropy energy
and to show that our method will yield convergent results for the whole
transition metal series and for large (Fe) and small (Ni) exchange
coupling.
We use parameters for freestanding Fe and Ni
monolayers and a lattice constant of 2.56~{\AA}
to simulate epitaxial growth on Cu(001).
In particular, our results yield that Fe has a
perpendicular easy axis with $E_{anis}=-0.32$~meV and Ni
an in-plane easy axis (along an axis connecting nearest neighbors)
with $E_{anis}=0.1$~meV~\cite{dipole}.
Experiment yields perpendicular anisotropy for ultrathin
Fe-films~\cite{expts}.
However, comparison to experiment for a monolayer is difficult
due to growth problems.
In the case of Ni, even the calculated in-plane direction of the easy axis
agrees with experiment~\cite{exptsni}.
Note, corresponding {\em ab initio} results for a free-standing Fe-monolayer
yielded $-0.4$~meV~\cite{wan93a,gay86},
but previous tight-binding calculations gave $-5.5$~meV~\cite{cin94}.
Although our numerical values of $E_{anis}$ depend on
the choice of parameters, the sign and the order of magnitude of
$E_{anis}$ are remarkably stable upon parameter variations:
In agreement with Wang {\em et al.}~\cite{wan93a} we find a
perpendicular easy axis also for Fe monolayers taking (001) surface
lattice constants of Pd, Ag, V, and W (2.77~{\AA},
2.89~{\AA}, 3.03~{\AA}, and 3.16~{\AA}, respectively).

It is of considerable interest for the physical interpretation of the
origin of magnetocrystalline anisotropy energy to study
$E_{anis}$ as a function of $\lambda_{so}$ and the hopping parameters $t$.
To a good approximation, we find quite generally
$E_{anis}\propto\lambda_{so}^2$, which sheds light on previous analyses.
This was calculated for Fe and Ni
bandstructure parameters for fixed bandfilling $n$
and in the range of $\lambda_{so}$ from zero
to the actual value (Fe: $\lambda_{so}=50$~meV, Ni: $\lambda_{so}=70$~meV).
However, for very small $E_{anis}$ the
dependence might be larger that $\lambda_{so}^2$.

Regarding the dependence of $E_{anis}$ on $\lambda_{so}$, the following
remarks are of interest. Treating SOC as a perturbation,
the lowest nonvanishing order of non-degenerate perturbation
theory is $\lambda_{so}^2$ due to time reversal symmetry~\cite{bru89}.
However, in the case of degenerate bands, a lifting of the degeneracy
by SOC will complicate the situation.
Degenerate bands can undergo a lifting linear in $\lambda_{so}$
when SOC is introduced, and the resulting contibution to $E_{anis}$ depends
on the area in {\bf k}-space influenced by the degeneracy.
Whether this area is of the order of $\lambda_{so}^2$,
which would yield $E_{anis}\propto\lambda_{so}^3$~\cite{wan93a},
or this area is of lower order and thus would yield an
important contribution to $E_{anis}$~\cite{kel93,pic94},
has been a controversial question.
This makes our result very interesting (see below).

Concerning the dependence of $E_{anis}$ on the $d$-electron hopping
parameters $t$, note that
the overall shape of the curves $E_{anis}(n)$
will not change if $t$ is varied. $|E_{anis}|$
increases for decreasing $t$ (decreasing bandwidth).
This leads to the general trend: $|E_{anis}|$ increases with
increasing lattice constant of the monolayer. This corresponds to an
increasing substrate lattice constant, if one neglects hybridization
effects between the substrate and the monolayer.

For the further physical interpretation of our results
for $E_{anis}$ as a function
of $n$ and $\lambda_{so}$, namely how $E_{anis}$ results
from the bandstructure and especially how it is
affected by the lifting of degeneracies in the bandstructure
close to $E_F$, we neglect $s$-$d$ hybridization for simplicity.
This is allowed since
only $d$-state degeneracies matter; $s$-$d$ hybridization changes only
the location of the degeneracies. The inset in
Fig.~\ref{X} shows schematically that a special type of degeneracy
(``line'' degeneracy) leads to an important contribution to
$E_{anis}$ if it occurs near $E_F$ for one direction of magnetization
and is lifted by SOC for another. Estimating this contribution (see
Fig.~\ref{X}) yields $\Delta E_{anis}=\frac{1}{2}
\lambda_{so}\cdot F$. The fraction $F$ of involved states in
{\bf k}-space is given by $F=\frac{\Delta k_1}{\frac{\pi}{a}}\cdot1=
2\lambda_{so}\left(\frac{\pi}{a}\frac{\partial E}{\partial
  k_1}\right)^{-1}$, since the intersecting bands are non-dispersive along one
direction in {\bf k}-space (perpendicular to ${\bf k}_1$).
Here, $a$ is the lattice constant, and
$\frac{\partial E}{\partial k_1}$ is the dispersion of the intersecting
bands near their intersection.
Hence, the maximum energy gained by the lifting of the degeneracy amounts to
\begin{equation}
\label{deltae}
\Delta E_{anis}=\lambda_{so}^2\left(\frac{\pi}{a}
                \frac{\partial E}{\partial k_1}\right)^{-1}.
\end{equation}
It is important to note that such contributions are not
caused by degeneracies at isolated {\em points}~\cite{degeneracies},
but by degeneracies along {\em lines} in ${\bf k}$-space.
Our bandstructure calculations yield such degeneracies
near $E_F$ for $n=8.7$ using Fe parameters
for the bandstructure and for $n=5.9$ using Ni parameters.

Eq.~(\ref{deltae}) explains immediately the height of the
peak at $n=8.7$ shown in Fig.~\ref{fig1} (curve a).
Taking from our bandstructure calculations (Fe parameters)
$\frac{\partial E}{\partial k_y}\approx1$~eV$/\frac{\pi}{a}$
we obtain $\Delta E_{anis}\approx2.5$~meV.
This is in excellent agreement with the exact result of our calculation
$E_{anis}(n=8.7)=2.7$~meV.
In addition, a detailed
{\bf k}-space resolved calculation of $E_{anis}(n=8.7)$ shows that
indeed only states near the corresponding degeneracy at $E_F$ yield
large contributions to $E_{anis}$.
Furthermore, Eq.~(\ref{deltae}) shows that
$\Delta E_{anis}\propto\frac{1}{t}$
since approximately $\frac{\partial E}{\partial k}\propto t$. Note, this is in
agreement with our previous observation regarding the dependence of $E_{anis}$
on the hopping parameters $t$.

Since the degeneracies are so important for $E_{anis}$, we outline in
the following at some length the
relationship between $E_{anis}$ and the
degeneracies and thus the crystal symmetry.
For that purpose, we adapt our basis set of
Bloch wave functions $\psi_i$, $i=1, ...,10$, in order to obtain a maximum
number of zeros in the Hamiltonian matrix for the $d$-electrons.
Expressed in terms of Bloch functions constructed from the
cartesian atomic orbitals, the $5{\times}5$
tight-binding matrix has its simplest block diagonal form with only two
off-diagonal elements (ODEs) $(H_{d})_{4,5}=(H_{d})_{5,4}$
(and equivalently $(H_{d})_{9,10}=(H_{d})_{10,9}$)
if the $x$-axis is directed along an axis connecting nearest neighbors.
To find out which additional
ODEs are introduced by SOC for a given direction of the magnetizaton
${\bf M}$ (in the following, ${\bf M}\parallel\hat{z}$ and ${\bf M}\parallel
\hat{x}$ are considered),
we analyze the form of $H_{so}$ in Eq.~(\ref{hso}).
We assume that different Bloch-states are orthogonal.
States with
parallel spins are coupled, if they contain equal orbital momenta with
respect to the spin quantization axis $\hat{\zeta}$, whereas states with
opposite spins must show a difference of one in the orbital momenta to yield
nonvanishing ODEs.
The cartesian orbitals $xy$, $yz$, $zx$, $x^2-y^2$ and $3z^2-r^2$
(leading to $\psi_1,...,\psi_5$ together with $\left|\uparrow\right>$ and to
$\psi_6,...,\psi_{10}$ with $\left|\downarrow\right>$, respectively)
are composed of eigenstates of $l_z$ with the eigenvalues
\mbox{(-2,2)}, \mbox{(-1,1)},
(-1,1), (-2,2) and 0, respectively. In terms of eigenstates of $l_x$
one finds the eigenvalues \mbox{(-1,1)}, \mbox{(-2,2)}, \mbox{(-1,1)}, (-2,0,2)
and (-2,0,2),
respectively.
Hence, coupling exists, for ${\bf M}\parallel\hat{z}$, within the groups of
states $\psi_i$ with $i=1,4,5,7,8$ and with $i=2,3,6,9,10$, and,
in the case of ${\bf M}\parallel\hat{x}$, within the groups of
states $\psi_i$ with $i=2,4,5,6,8$ and $i=1,3,7,9,10$, respectively.
In both cases, the Hamiltonian can be split into two $5{\times}5$ blocks,
and subbands belonging to different blocks will intersect. Between states
of the same block, the degeneracies will ordinarily be removed.
Especially the subbands $\psi_1$ and $\psi_2$
(and, correspondingly, $\psi_6$ and $\psi_7$)
change their roles, if the magnetization is changed from $\hat{z}$
to $\hat{x}$ and vice versa, because the orbitals $xy$ and $yz$ have
different orbital momenta with respect to the $x$- and $z$-axes.
So, they will be involved in the lifting of
degeneracies by altering magnetization and possibly, as shown above, yield
important contributions to $E_{anis}$.
As an example, the prominent peak in
the $E_{anis}(n)$ curve of Fe at $n=8.7$  results from a degeneracy of
the subbands corresponding to the states $\psi_7$ and $(\psi_9,\psi_{10})$
for ${\bf M}\parallel\hat{z}$,
occurring along a line parallel to the $k_x$-axis in {\bf k}-space,
that is lifted
for ${\bf M}\parallel\hat{x}$, because in the second case the subbands
belong to the same block of the Hamiltonian, whereas in the first they do not.

Concerning the different behavior of $E_{anis}$ as
a function of $n$ choosing Fe or Ni parameters (see Fig.~\ref{fig1}),
one should note that due to
the large exchange splitting in Fe only the SOC between $d$-states with
{\em parallel} spins contributes significantly to $E_{anis}$.
Thus, $E_{anis}$ for a less than half-filled $d$-band ($n<6.2$
in Fig.~\ref{fig1}, curve a)
mainly results from the majority-spin subband and $E_{anis}$ for
a more than half-filled $d$-band ($n>6.2$) from
the minority-spin states. This explains the similar shape of these
two parts of the curve $E_{anis}(n)$.
For Ni, the contribution to $E_{anis}$ from SOC between opposite spin
states is of equal magnitude, leading to a rather different behavior
of $E_{anis}(n)$.

Apparently, the occurrence and lifting of
degenerate subbands for different directions of magnetization is important
for calculating the magnetic anisotropy energy.
This should also be clearly seen from the temperature dependence of
$E_{anis}$ due to the magnetization ${\bf M}(T)$,
the Fermi function $f_T(\Delta E)$,
the hopping integrals $t(T)$, which depend on temperature due to
lattice expansion, and the entropy $S(T)$.
In particular, effects resulting from the latter three contributions
are analyzed.
We find a characteristic energy scale of the order of magnitude of
$\lambda_{so}$ for the reduction of the free anisotropy energy
$F_{anis}=E_{anis}-TS_{anis}$ with increasing temperature.
This is demonstrated in Fig.~\ref{fig3} in a
$d$-band calculation with Fe parameters ($d$-bandfilling $n_d=6.0$), where
$F_{anis}$ decreases over
a temperature range of 500--1000~K (50--100~meV), which corresponds to
the energy
$\lambda_{so}$, but not to the $3d$-bandwidth of about 3~eV.
This becomes immediately plausible if one notices that the
SOC-induced
lifting of degeneracies occurs near the Fermi-level. Thus one
expects a measurable effect on $F_{anis}$ if $k_B T$ becomes larger than
$2\lambda_{so}$.
Furthermore we must conclude from our results that
shifting of subbands far below the Fermi-level is not so important, since
then $F_{anis}$ could not be essentially lowered on such a
small temperature scale.
Hence, this analysis of $F_{anis}(T)$ shows also the significant role
of changes of the
degeneracies giving an important contribution to the anisotropy energy.
It is remarkable that the three temperature effects mentioned above
are of equal magnitude as the temperature effects due to ${\bf M}(T)$.

In conclusion, a fully convergent calculation of the
magnetocrystalline anisotropy energy $E_{anis}$ of Fe and Ni
monolayers on Cu (001) is performed. We find a perpendicular easy axis
for Fe and an in-plane easy axis for Ni.
Large contributions to $E_{anis}$
result from the SOC-induced lifting of degeneracies at the
Fermi-level, which is also supported by our calculation of
temperature effects. In general, $E_{anis}$ scales with the square of
the SOC constant $\lambda_{so}$.
The main features of our analysis can be applied also to
thicker films, which is of particular interest
for the investigation of reorientation
transitions.

\newpage
\noindent
\begin{figure}
\caption{Dependence of $E_{anis}$ on the $3d$- and $4s-$bandfilling $n$
for a monolayer
with parameters referring to Fe (curve a) and Ni (curve b).
Negative values of $E_{anis}$ yield perpendicular anisotropy. The vertical
lines denote $n$ for Fe and Ni, respectively.}
\label{fig1}
\end{figure}
\begin{figure}
\caption{
Temperature dependence of $F_{anis}(T)$ for a Fe-parametrized
$d$-band calculation with $d$-bandfilling $n_d=6$.
The inset shows the
occurrence (full lines) and lifting (dashed lines) of a ``line''
degeneracy for two different
directions of magnetization ${\bf M}_1$ and ${\bf M}_2$,
respectively.
${\bf k}_1$ corresponds to one particular direction in {\bf
k}-space. Perpendicular to ${\bf k}_1$ the intersecting bands are
non-dispersive throughout the BZ.
Note, the energy gained by the
lifting of this degeneracy is given by $\Delta E_{anis}=\frac{1}{2}
\lambda_{so}\cdot F$, if $E_F$ falls in between the two subbands
(dotted line). Here, $F$ is the fraction of the involved states in
{\bf k}-space. Apparently, if $E_F$ lies
below or above the two subbands, $\Delta E_{anis}$ is zero.}
\label{X}
\label{fig3}
\end{figure}

\begin{thebibliography}{99}
\bibitem {angles} $\theta$ denotes the angle between the magnetization
  direction $\hat{\zeta}$ and the surface normal $\hat{z}$.
The in-plane angle $\phi_0$ is chosen such that the resulting $|E_{anis}|$
is the largest possible.
\bibitem {dipole} We obtain for the classical dipole-dipole interaction energy
  $E_{dip}=\frac{1}{2}\mu_0 M^2=0.2$~meV per Fe-atom. Here, the
  magnetization $M$ results
  from the magnetic moment $m=3.2\mu_B$ per atom.
  Including $E_{dip}$, we still find a
  perpendicular easy axis for Fe on Cu(001). For Ni, we obtain
  $E_{dip}=0.02$~meV from $m=0.91\mu_B$. This confirmes the in-plane easy axis.
\end{thebibliography}
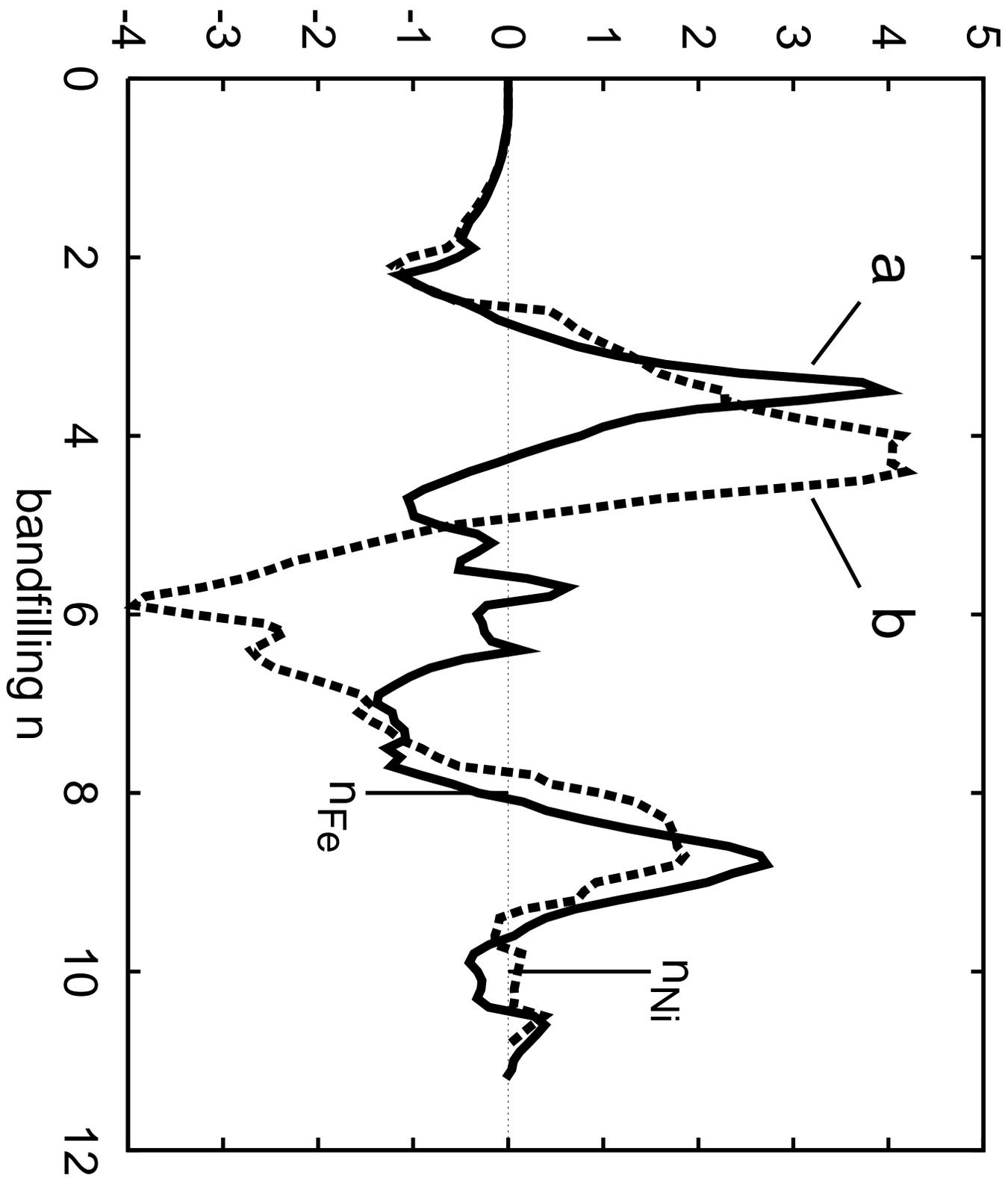
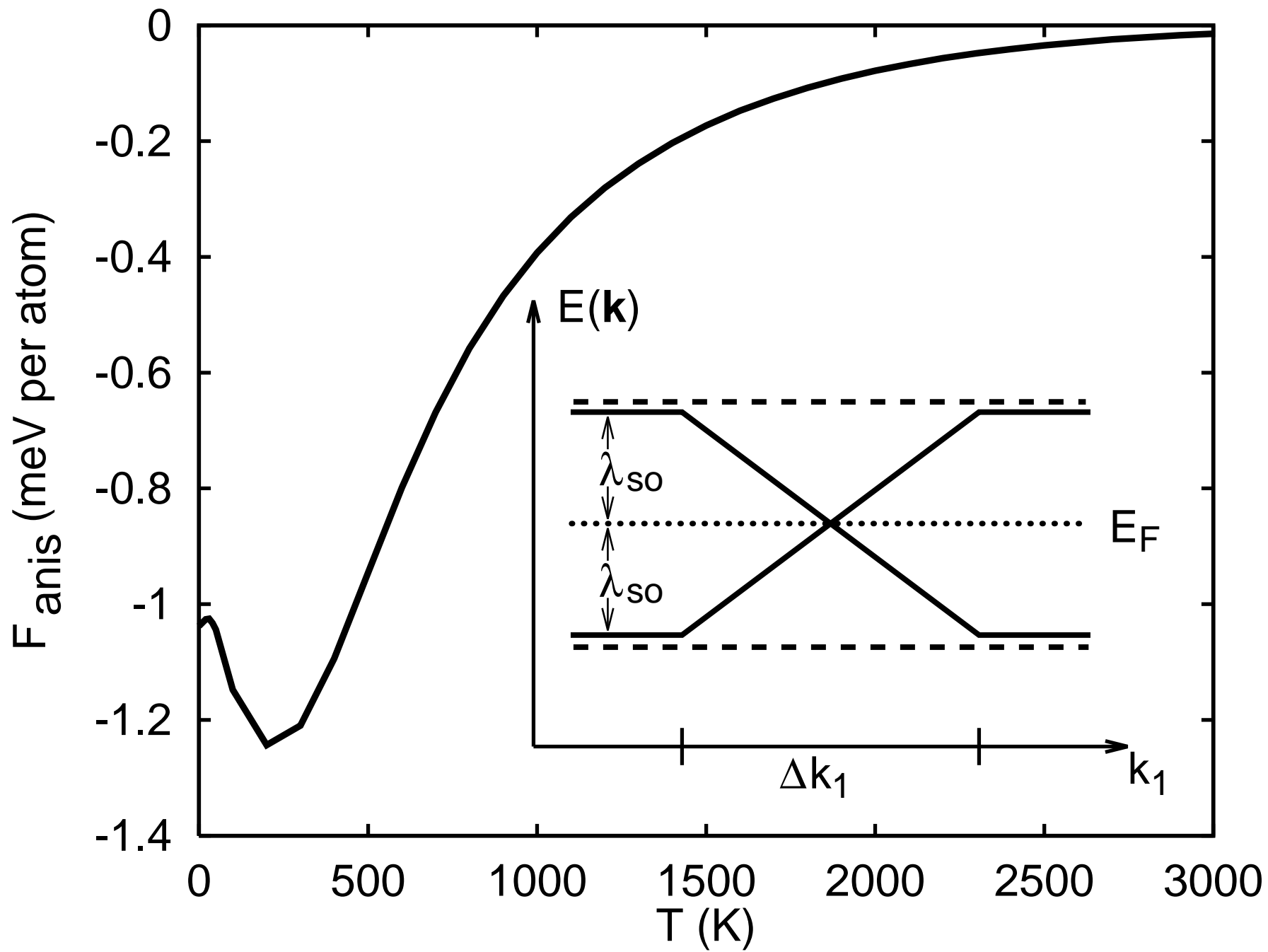
\end{document}